\documentclass[letter]{ptptex}

\usepackage{amssymb}
\usepackage[dvips]{graphicx}
\usepackage{amsmath,amscd}
\usepackage{color}

\def\a{\alpha}
\def\a'{\alpha '}

\def\ep{\epsilon}

\def\Lam{\Lambda}

\def\L{\mathcal{L}}

\def\N{\mathcal{N}}

\def\Tr{{\rm Tr}}

\def\1{\amssymb{1}}

\def\lb {\left(}
\def\rb {\right)}

\def\lan{\langle}
\def\ran{\rangle}

\newcommand {\beq}{\begin{eqnarray}}
\newcommand {\eeq}{\end{eqnarray}}
\newcommand {\ben}{\begin{enumerate}}
\newcommand {\een}{\end{enumerate}}
\newcommand {\bit}{\begin{itemize}}
\newcommand {\eit}{\end{itemize}}

\makeatletter
\@addtoreset{equation}{section}

\makeatother

\makeatletter
\newcommand{\thetablename}{Table}
\def\fnum@table{\thetablename\ \thetable}
\makeatother
\begin{document}

\thispagestyle{empty}
\begin{flushright}

{\tt arXiv:0903.3637} \\
March, 2009\\
\end{flushright}
\vspace{3mm}

%%%%%%%%%%%%%%%%%%%%%%%%%%%%%%%%%%%%%%%
\begin{center}
{\huge Higgs\,-\,Inflaton Potential\\
  \vspace{2mm}in Higher\,-\,Dimensional SUSY Gauge Theories}

\end{center}
%%%%%%%%%%%%%%%%%%%%%%%%%%%%%%%%%%%%%%%%%

\begin{center}
\lineskip .45em
\vskip1.5cm
{\large Takeo Inami\footnote{E-mail: inami@phys.chuo-u.ac.jp}$^1$,
Yoji Koyama\footnote{E-mail: koyama@phys.chuo-u.ac.jp}$^1$, C. S. Lim\footnote{E-mail: lim@kobe-u.ac.jp}$^2$ and
Shie Minakami\footnote{E-mail: minakami@phys.chuo-u.ac.jp}}$^1$

\vskip 1.5em
{\large\itshape $^1$Department of Physics, Chuo University, Bunkyo-ku, Tokyo 112, Japan\\
\large\itshape $^2$Department of Physics, Kobe University, Nada, Kobe 657, Japan}  \vskip 4.5em
\end{center}

\begin{abstract}
\vspace{2mm}
We study the possibility that the Higgs and the inflaton are the same single field or cousins arising from the extra space components of some higher-dimensional gauge field. We take 5D supersymmetric gauge theory with a matter compactified on $S^1$ as a toy model and evaluate the one-loop contribution to the Higgs-inflaton potential. Our gauge-Higgs-inflaton unification picture applied to the gauge field of intermediate energy scale ($\sim 10^{13}$ GeV) can explain the observed inflation parameters without fine-tuning.   
\end{abstract}

%%%%%%%%%%%%%%%%%%%%%%%%%%%%%%%%%%%%%%%%%%%%%%%%%%%%%%

%%%%%%%%%%%%%%%%%%%%%%%%   1   %%%%%%%%%%%%%%%%%%%%%%%%%%
\newpage

\section{Introduction}

Evaluation of the Higgs potential in the unified gauge theory encounters the fine-tuning problem in the coupling constants, so called gauge hierarchy problem. There have been a few alternative solutions to this problem, technicolor model, supersymmetry (SUSY) and later higher-dimensional gauge theory \cite{Hatanaka:1998yp}. Recently, people began to see that there is an analogous fine-tuning problem in the evaluation of the inflaton potential \cite{ArkaniHamed:2003wu,Kaplan:2003aj}, when they try to explain the cosmological inflation parameters. Most models so far proposed have difficulty in explaining this fine-tuning in a natural way as pointed out in \cite{ArkaniHamed:2003wu}.\\ 
\indent Assuming that the 4D gauge theory is an effective field theory of the superstring theory of some sort, there are only few origins of scalar fields. One is the extra space components of higher-dimensional gauge fields in open string. We pursue the possibility that the two fine-tuning problems, one in the Higgs potential and the other in the inflaton potential, are related and may be solved by assuming that the two scalar fields have the same origin, i.e. the extra space components of gauge fields in higher dimensional theory.\\ 
\indent To study this view we take 5D $\N=1$ supersymmetric $SU(2)$ gauge theory with a matter multiplet compactifed on $S^1$ as a toy model, and denote the gauge field by $A_M$ $(M=0,\cdots,3,5)$. We identify the zero mode of its 5th component $A_5^{(0)}$ as the Higgs and the inflaton $\phi$. Here we take $SU(2)$ as the simplest non-Abelian gauge group; the extension to other gauge groups should be easy. To have SUSY breaking in 5D supersymmetric theory the Scherk-Schwarz mechanism \cite{Scherk:1978ta} is the most economical one, which we will use and we evaluate the one-loop contribution to the Higgs-inflaton potential. This way we also study the role of supersymmetry breaking in the two fine tuning problems. Arkani-Hamed et al. already pointed out the reliable perturbative computation of the potential in higher-dimensional gauge theories\cite{ArkaniHamed:2003wu}. The same remark should apply to supersymmetric models as well.\\
\indent The parameters of the model are fixed such that the conditions for inflation are met. The gauge coupling constant $g$ turns out to be as large as 0.9, being close to the realistic value of the gauge coupling constant. This gratifying result may be by chance but looks encouraging to us. 

%%%%%%%%%%%%%%%%%%%%%%  2  %%%%%%%%%%%%%%%%%%%%%%%%%%%

\section{The Model and One-Loop Effective Potential}

We consider 5D $\N=1\,SU(2)$ super Yang-Mills (SYM) theory compactified on $S^1$. In five dimension, the vector multiplet consists of the gauge field $A_M \,(M=1,\cdots,5)$, symplectic-Majorana spinors $\lambda_L^i\, (i=1,2)$ and a real scalar $\Sigma$. A hypermultiplet matter consists of complex scalar fields $H_i$ and a Dirac fermion $\Psi=(\Psi_L,\ \Psi_R)^T$. $\lambda_L^i$ and $H_i$ are doublets of the $SU(2)_R$ symmetry. The Lagrangian is given by the sum of two terms \cite{Pomarol:1998sd},
\beq
\L^{\rm{gauge}}=\Tr\left[-\frac{1}{2}(F_{MN})^2+(D_{M}\Sigma)^2+i\bar{\lambda_{i}}\gamma^MD_{M}\lambda^i-\bar{\lambda_{i}}[\Sigma,\lambda^i]\right],
\eeq
\beq
\L^{\rm{matter}}&=&|D_{M}H_{i}|^2-m^2|H_{i}|^2+\bar{\Psi}(i\gamma^MD_{M}-m)\Psi-(i\sqrt{2}g_{5}H^{\dag i}\bar{\lambda_{i}}\Psi+h.c)\nonumber\\
&&-g_{5}\bar{\Psi}\Sigma\Psi-g_{5}mH^{\dag i}\Sigma H_i-g_{5}^2H^{\dag i}\Sigma^2H_i-\frac{g_{5}^2}{2}\sum_{a,A}[H^{\dag i}(\sigma^a)^j_iT^AH_j]^2,
\eeq
where $m$ is the matter mass and $g_{5}$ is the 5D gauge coupling constant. The covariant derivative acting on the fundamental representation is defined as $D_MH^i \equiv \partial_MH^i-ig_5A_MH^i$. $\sigma^a\,(a=1,2,3)$ and $T^A\,(A=1,2,3)$ are Pauli matrices belonging to the $SU(2)_R$ and the gauge group $SU(2)$, respectively.\\ 
\indent After compactification on $S^1$, the 5th component $A^{(0)}_5$ of the zero modes $A^{(0)}_M$ is a 4D scalar. We denote the 5D coordinate by $x^M=(x^{\mu},y)$. We identify $A^{(0)}_5$ with the inflaton field $\phi$, as will be discussed later. To have nonzero effective potential for $A_5^{(0)}$, SUSY has to be broken. A few ways of SUSY breaking are known. A natural way in the 5D theory is the Scherk-Schwarz mechanism \cite{Scherk:1978ta} associated with $SU(2)_R$ symmetry. We employ twisted boundary conditions for $\lambda^i$ and $H^i$ in the $S^1$ direction $y$.
\beq
\lambda_i(x^{\mu},y+2\pi R)=(e^{i\beta^a\sigma^a})_i^{\ j}\lambda_j(x^{\mu},y)\quad\mbox{same for }H^{\dag i}.
\eeq
$R\,(=L/2\pi)$ is the $S^1$ radius and $\beta^a$ is the SUSY breaking parameter. We will later use $\beta=\sqrt{(\beta^a)^2}$.\\ 
\indent To evaluate the effective potential, we allow $A_5^{(0)}$ to have VEV of the form 
\beq
\lan A_5^{(0)}\ran= \frac{1}{g_5L}\left(\begin{array}{cc}
\theta& 0\\
0 & -\theta
\end{array}\right).
\eeq
$\theta$ is a constant given by the Wilson line phase, ${\rm diag}\,(\theta, -\theta)=g_5\int_0^Ldy\lan A_5^{(0)}\ran$. We consistently set the VEV of the scalar fields $\Sigma$ and $H^i$ to zero. 
We study the effective potential for $\theta,\,\,V(\theta)$. The one-loop terms can be easily computed in the background field method. The result is obtained in the same way as \cite{Delgado:1998qr}
\beq
V^{\rm{gauge}}(\theta)=-\frac{8\Gamma(\frac52)}{\pi^{5/2}}\frac{1}{L^4}\sum_{n=1}^{\infty}\frac{1}{n^5}
(1+\cos(2n\theta))(1-\cos(n\beta))+ \mbox{const}\label{gaugep},
\eeq 
\beq
V^{\rm{matter}}(\theta)&=&\frac{2\sqrt{2}}{\pi^{5/2}}\sum_{n=1}^{\infty}\left(\frac{m}{nL}\right)^{\frac52}\left(\frac{\pi L}{2mn}\right)^{\frac12}\left(1+\frac{3}{mLn}+\frac{3}{(mLn)^2}\right)\nonumber\\
&& \times e^{-mLn}\cos(n\theta)(1-\cos(n\beta))+ \mbox{const}\label{hyperp}.
\eeq
There are divergent constant terms, which we will later renormalize to the appropriate value. $V^{\rm{gauge}+\rm{matter}}(\theta)$ has minima at $\theta=\pi +2\pi k$ and maxima at $\theta=\pi/2+2\pi k\,(k=\mbox{integer})$.\\
\begin{figure}[tbp]
 \begin{center}
  \includegraphics[width=100mm]{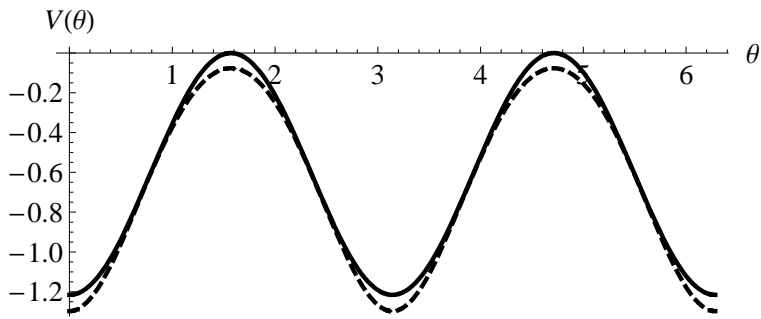}
\vskip-\lastskip
\caption{The effective potential $V^{\rm{gauge}}$ with $\beta=\pi/2$. The potential evaluated by taking the summation up to $n=100$ is represented by a broken line, the potential evaluated by taking only the $n=1$ term by a solid line. The difference between the two is small. Here the unit of the longitudinal axis is $1/L^4$. The same convention is used in Fig. 2 and 3. }
 \label{fig:one}
\end{center}
\end{figure}
\indent The leading term ($n=1$) is already a good approximation in (\ref{gaugep}) (see Fig. 1) and (\ref{hyperp}).
\beq
V^{\rm{gauge}}(\theta)\simeq-\frac{6}{\pi^{2}}\frac{1}{L^4}
(1+\cos(2\theta))(1-\cos \beta)+ \mbox{const}\label{gaugeap},
\eeq
\beq
V^{\rm{matter}}(\theta)\simeq \frac{2}{\pi^{2}}\frac{1}{L^4}\left((mL)^2+3{mL}+3\right) e^{-mL}\cos \theta(1-\cos \beta)+ \mbox{const}\label{hyperap}.
\eeq
The gauge and matter parts have the same $\beta$ dependence, but have different periodicity in $\theta$. They have different sign for $\theta \leq \pi/2$ and the same sign for $\pi/2\leq \theta$. They tend to cancel each other for $\theta \leq \pi/2$. A heavy hypermultiplet ($m\gg 1/L$) does not contribute to $V(\theta)$ due to factor $e^{-mL}$. \\
\indent From now on we consider the theory near one of the potential minima $\theta_0$, and set
\beq
\theta(x^{\mu})=\theta_0+\omega(x^{\mu}).
\eeq
The 4D effective Lagrangian for the field $\omega$ is
\beq
\L=\frac{1}{2}f^2(\partial_{\mu}\omega)^2-V^{\rm{gauge}+\rm{matter}}(\omega+\theta_0)\label{Lag}
\eeq
 Here $f$ is related to the Higgs-inflaton mass (see (\ref{infmass})), 
\beq
f=\frac{1}{2\pi gR},\label{ssbs}
\eeq
where $g=g_5/\sqrt{L}$ is the 4D gauge coupling constant.

%%%%%%%%%%%%%%%%%%%%%%%  3  %%%%%%%%%%%%%%%%%%%%%%%%%%%

\section{Gauge Hierarchy and Inflation}

We consider the possibility that the Higgs and the inflaton are the same field $\phi$ and identify them with $A^{(0)}_5$. We study the question of whether the fine-tuning problem in the inflaton potential may be solved in this view. We set 
\beq
\phi=\omega f. \label{gf}
\eeq 
We begin by recapitulating the implication for the inflaton potential from the recent astrophysical data \cite{Lyth:1998xn,Komatsu:2008hk}.
\ben

\item[1)] We choose the renormalization of the effective potential $V$ so that it satisfies, $V\vert_{\mbox{min}}=0$, in accordance with the nearly zero cosmological constant $\Lam \simeq 10^{-12}\,({\rm eV})^4$.

\item[2)] The slow-roll conditions 
\beq
\epsilon \equiv \frac{1}{2}M_{P}^2\lb\frac{V'}{V}\rb^2\ll1, \quad \eta \equiv M_{P}^2\frac{V''}{V}\ll1,\label{slow-roll}
\eeq
where $M_P = 2.4\times10^{18}$ GeV is the reduced Planck mass, and $V'=dV/d\phi$.

\item[3)] The spectral index $n_s$
\beq
n_s \equiv 1-6\epsilon+2\eta, \quad 0.948 \leq n_s \leq 0.977\label{spectral},
\eeq
taking account of the latest five-year WMAP deta \cite{Komatsu:2008hk}. This condition is satisfied if $\epsilon$, $\eta \sim 0.01$. 

 \item[4)] The number of e-foldings $N$ 
\beq
N \equiv \int_{t_i}^{t_f} Hdt \simeq \frac{1}{M_P^2} \left| \int_{\phi_i}^{\phi_f} \frac{V}{V'} \,d\phi \right|.\label{N}
\eeq
The second equality holds true under the slow-roll condition. For solving the flatness and the horizon problems the observable inflation\footnote{Inflation can be investigated by observations only after the observable Universe has left the horizon. This era of inflation is called the observable inflation.} must occur for a sufficiently long time, namely $N$ has to be 50 - 60. The Hubble parameter $H$ is integrated over the observable inflation period; $t_i$ is the time when the observable Universe leaves the horizon and is determined by the conditions (\ref{slow-roll}) and (\ref{N}). $t_f$ is the time when the slow-roll condition ends, namely when $\ep$ and $\eta$ become nearly 1.
  $\phi_i$ and $\phi_f$ are the values of $\phi$ at $t_i$ and $t_f$. 
\item[5)] The curvature perturbation $\delta_H$ is written in terms of $V$ (see for instance \cite{Lyth:1998xn}). Substituting its experimental value \cite{Komatsu:2008hk} to $\delta_H$, 
\beq
\delta_H=\frac{1}{5\sqrt{6}\pi}\frac{V^{1/2}}{M_{P}^2\epsilon^{1/2}} =1.91\times 10^{-5}.\label{delta}
\eeq

\item[6)] Quantum gravity effects can be neglected if the $S^1$ compactification scale $L$ is larger than  the Planck length $1/M_{P}$ \cite{Appelquist:1983vs},
\beq
R \gtrsim 3\cdot (16\pi^{2}M_{P})^{-1}\simeq 0.78\times10^{-20}\,\mbox{GeV}^{-1}.
\label{qgmusi}\eeq
\een
\indent $V$ appearing in (\ref{delta}) and (\ref{spectral}) is understood to be its value at $\phi=\phi_*$, the value of $\phi$ related to the epoch $t_*$ of horizon exit for a certain length scale, and in practice $\phi_* \simeq \phi_i$. See \cite{Lyth:1998xn} for the precise definition of $\phi_*$.\\
\indent In our 5D model the inflaton is the Higgs $h$ itself, hence $m_{\phi}^2=m_{h}^2$. We wish to see whether the two fine-tuning problems can be solved simultaneously in our view. In applying our idea to unified theories, we will see that the $m_{\phi}$ is fixed by the value of $\delta_H$ 
through (\ref{delta}).\\
 \indent We study whether the conditions for inflation summarized above are met for reasonable values of $g,\ R\ \mbox{and}\ \beta$. We consider two 5D supersymmetric models, (i) model of pure SYM and (ii) model of SYM + one hypermultiplet. We take the $n=1$ approximation (\ref{gaugeap}) and (\ref{hyperap}) to $V(\phi/f)$. For the case (ii), the size of $V^{\rm{matter}}$ changes slightly for different values of matter mass $m$ in the range $m\lesssim2/L$. We will take $m=1/L$ as its typical value. We will see that the two models yield similar results.\\ 
\indent Before proceeding to the analysis of our model we note that the potential $V$ as a function of $\phi$ differs depending on the value of $f$. The potential in the model of pure SYM is shown for two typical values of $f$, $50M_P$ (broken line) and $10M_P$ (solid line) in Fig 2. 
\begin{figure}[tbp]
 \begin{center}
  \includegraphics[width=100mm]{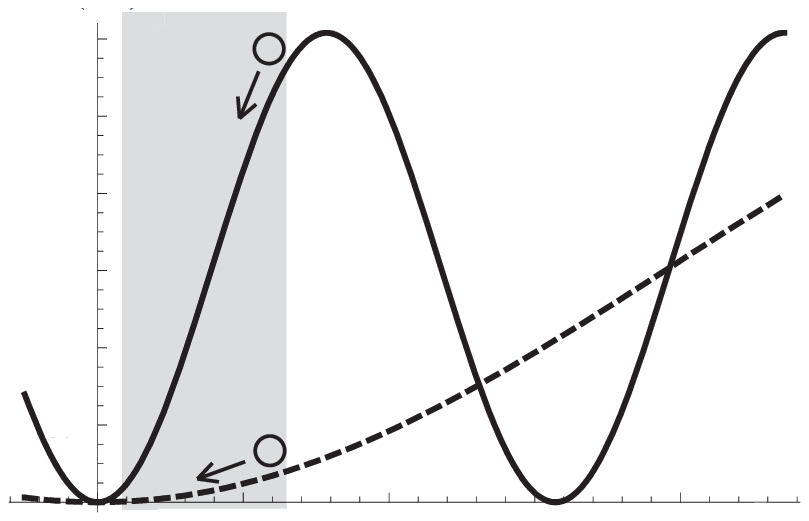}
\put(-212,6){$10$} \put(-164,6){$20$}\put(-119,6){$30$}\put(-73,6){$40$}\put(-34,28){$\phi$}\put(-27,6){$\times M_P$}
\put(-272,39){$0.2$}\put(-272,64){$0.4$}\put(-272,89){$0.6$}\put(-272,113){$0.8$}\put(-272,137){$1.0$}\put(-272,162){$1.2$}\put(-272,180){$V(\phi/f)$}
 \end{center}
\vskip-\lastskip
\caption{The inflaton potential in the case (i) for two values of $f$. The shaded area represents the region of observable inflation.}
 \label{fig:two}
\end{figure}
The region of observable inflation, $\phi_f \leq \phi \leq \phi_i$, is shown as the shaded area in the latter case. $\phi_i$ is determined from the condition (\ref{N}), and it is $16M_P$ and $13M_P$ in the two cases. Note that $V\sim \phi^2$ is a good approximation for the dashed line whereas the $V\sim 1-\cos(2\phi/f)$ should be used for the solid line.
 \begin{flushleft}
{\bf (i) pure SYM}
\end{flushleft}
In accordance with the condition 1) we set
\beq
 V(\phi/f)&\equiv& V^{\rm{gauge}}(\phi/f)-V^{\rm{gauge}}(0)\nonumber \\
&=&  \frac{6}{\pi^{2}}\frac{1}{L^4}
(1-\cos(2\phi/f))(1-\cos\beta).\label{potential}
\eeq
$\epsilon$ and $\eta$ are given in terms of $\phi$ as
\beq
\epsilon = 2\lb\frac{M_{P}}{f}\rb^2\frac{1+\cos(2\phi/f)}{1-\cos(2\phi/f)}, \quad \eta =4\lb\frac{M_{P}}{f}\rb^2\frac{\cos(2\phi/f)}{1-\cos(2\phi/f)}.\label{slow-rollpara}
\eeq
The conditions 2) - 4) are met if $f^2 \gg M_P^2$ ($f\gtrsim10M_P=2.4\times 10^{19}$ GeV), and hence
\beq
gR\lesssim 6.6\times 10^{-21}\,\mbox{GeV}^{-1}\label{fMP}.
\eeq
The condition 5) from $\delta_H$ gives a relation between $R$ and $\beta$. Then the equality is satisfied if 
\beq
R^2=4.7\times\sqrt{1-\cos \beta}\times\left(\frac{1-\cos(2\phi_*/f)}{\epsilon}\right)^{1/2}\times 10^{-36}\,\mbox{GeV}^{-2}\label{deltaH}.
\eeq
The values of $R$ are constrained by (\ref{qgmusi}) and (\ref{deltaH}).
\beq
0.8\times 10^{-20}\,\mbox{GeV}^{-1}\lesssim R = 2.2\times (1-\cos \beta)^{1/4}\times \left(\frac{1-\cos(2\phi_*/f)}{\epsilon}\right)^{1/4} \times 10^{-18}\,\mbox{GeV}^{-1}.\nonumber \\ \label{rangeR}
\eeq
Note that the lower bound is from a theoretical consideration whereas 
the equality  on the right side is derived from the data on $\delta_H$. 
Rather small values of $\beta$ are implied in order that the inequality (\ref{rangeR}) makes sense.
\beq
\cos \beta\lesssim 1-1.7\times  \left( \frac{\epsilon}{1-\cos(2\phi_*/f)}\right)\times10^{-10}.\label{beta}
\eeq
\indent The allowed values of $g$ are found by combining (\ref{qgmusi}) and (\ref{fMP}).
\beq
g \lesssim 0.85 \label{g}.
\eeq 
The parameters $g$, $R$ and $\beta$ are further related to each other so that 
the inflaton mass $m_{\phi}$ is reproduced. We have from (\ref{potential}) 
\beq
m_{\phi}^2 = 6 \frac {1 - \cos \beta} {\pi^4} \frac {g^2}{R^2}.\label{infmass}
\eeq
\indent The Higgs-inflaton potential we have constructed from higher-dimensional gauge theories is meant to be a toy model. Nevertheless it is curious to see whether this simple inflaton model has some 
realistic meaning in the context of GUT or other unified models.\\
\indent We first recall that the $m_{\phi}$ 
is constrained severely by the value of $\delta_H$ through (\ref{delta}). 
The $m_{\phi}$ turns out to differ a bit depending on 
whether we take the quadratic function or the exact cosine 
function for the potential $V$. For large values of $f$ ($f \gg 10 \, M_P$) the potential 
(\ref{potential}) is approximated by the quadratic term (case II). Then the model is 
reduced to the time-honored chaotic potential. In this case the known result can be used, $m_{\phi} = 1.8 \times 10^{13}$ GeV \cite{Lyth:1998xn}. For $f \simeq \, 10M_P$ the potential cannot be approximated by the quadratic term; we have to evaluate $V(\phi/f)$ using the cosine 
function at $\phi_*$ (case I). The value $V(\phi_*/f)$ determines the inflaton mass through the condition 5). We have from (\ref{deltaH}) and $\phi_*\simeq12M_P$
\beq
\frac{1}{R^2}\sqrt{1-\cos \beta}=8.9\times10^{33}\,{\rm GeV}^2.\label{cosmass}
\eeq
Using the equality of (\ref{fMP}), (\ref{infmass}) and (\ref{cosmass}), we get $m_{\phi} = 1.3 \times 
10^{13}$ GeV. The results in both case imply that our Higgs-inflaton may be  connected to some intermediate symmetry breaking in GUT, such as $SO(10)$ model.\\
\indent We now apply our Higgs-inflaton model to gauge theories for two typical values of $f$.\\
I ) $f=10M_P$. The potential $V(\phi_*/f)$ should be treated as the cosine 
function. Using $m_{\phi} \sim 1.3 \times 10^{13}$ GeV, we have 
\beq
6 \frac {1 - \cos \beta} {\pi^4} \frac {g^2}{R^2}\simeq 1.7\times10^{26}\,{\rm GeV}^2.\label{iGUT1}
\eeq
We estimate the values of the parameters by using (\ref{rangeR}) through (\ref{infmass}) and (\ref{iGUT1}).
\beq
g \simeq 0.9\,-\, 0.0005,\quad
R\simeq (8\times10^{-21}\,-\,1\times10^{-17})\, \mbox{GeV}^{-1},\quad
\beta \simeq 8\times10^{-7}\,-\,3.\label{range}
\eeq
II ) $f=100M_P$. The potential $V(\phi_*/f)$ can be approximated by the quadratic 
term. The same argument as (\ref{potential}) through (\ref{infmass}) applies to this case. 
Using $m_{\phi} \sim 1.8 \times 10^{13}$ GeV, we find the following values 
of the parameters.
\beq
g \simeq 0.09\,-\, 0.0002,\quad
R\simeq (8\times10^{-21}\,-\,3\times10^{-18})\, \mbox{GeV}^{-1},\quad
\beta \simeq 1\times10^{-5}\,-\,3.\label{rangeII}
\eeq
\indent It is of interest to compare the 4D gauge coupling constant 
$g$ with the realistic value, $g_{\rm{GUT}} = 0.7$ in the SUSY $SU(5)$ GUT. Curiously, in the case I the upper value of $g$ has turned out to be close to $g_{\rm{GUT}}$. On the other hand, in the case II the value of $g$ is one tenth of $g_{\rm{GUT}}$.
\begin{flushleft}
{\bf (ii) SYM + a hypermultiplet}
\end{flushleft}
The sum of the gauge part and the matter part, 
$V(\theta) = V^{\rm{gauge}} + V^{\rm{matter}}$, 
has the periodicity 2$\pi$ as shown in Fig. 3.
\begin{figure}[tbp]
 \begin{center}
  \includegraphics[width=100mm]{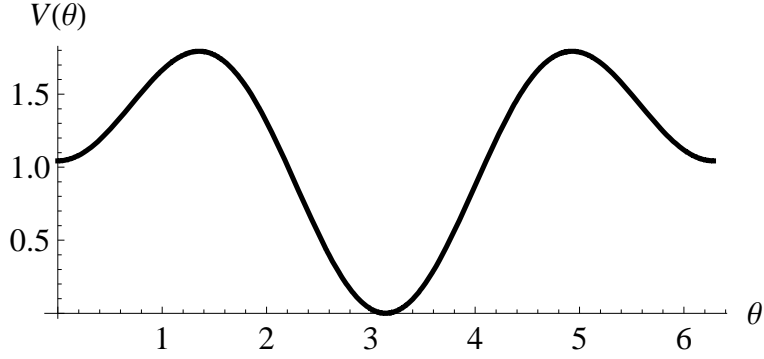}
 \end{center}
\vskip-\lastskip
\caption{The effective potential $V(\theta)=V^{\rm{gauge}}+V^{\rm{matter}}$ with $\beta=\pi/2$.}
 \label{fig:three}
\end{figure}
It has a minimum 
at two points, $\theta_0$ = 0 and $\pi$. The latter is the true minimum.  
In Fig. 3, taking account of the condition 1), we have set 
\beq
 V(\phi/f)&\equiv& V^{\rm{gauge}+\rm{matter}}(\phi/f+\pi)-V^{\rm{gauge}+\rm{matter}}(\pi)\nonumber \\
 &=& \left[6(1-\cos(2\phi/f))+14e^{-1}(1-\cos (\phi/f))\right]\frac{1}{\pi^2 L^4}(1-\cos\beta). \label{potential2}
\eeq
The inflaton mass is now given by 
\beq
m_{\phi}^2=7.3\frac{(1-\cos\beta)}{\pi^4}\frac{g^2}{R^2}.\label{imass}
\eeq
Note that the numerical coefficient is now 7.3 in place of 6 in (\ref{infmass}) 
in the model of pure SYM. The main difference between the model 
of SYM + matter and the pure SYM model comes from the value of $V(\phi_*/f)$. The effect of including a matter amounts to the change of $V(\phi_*/f)$ by a factor less than 2. As a result, the estimate of the 
parameters $g$, $R$ and $\beta$ proceeds in the same way as in (i). 
We obtain almost the same results as (\ref{range}) and (\ref{rangeII}) for the cases I and II. 
The previous comments on the physical meaning of our model applied 
to the Higgs of intermediate energy scale in GUT hold true. 
%%%%%%%%%%%%%%%%%%%%%%%  4  %%%%%%%%%%%%%%%%%%%%%%%%%%%%
\section{Discussion}
To construct an inflation model which meets the condition of slow-roll and tiny curvature perturbation without fine tuning the potential is an urgent problem in early cosmology. In higher-dimensional gauge theory small (in size) and flat potential of scalar fields arises naturally through loop corrections. It is interesting to see if the two fine-tuning problems, one in the inflaton model and the other in gauge symmetry breaking, are solved simultaneously in this view. We have studied the one-loop inflaton potential in 5D SYM theory as a toy model for this mechanism. \\
\indent An inflaton potential with the desirable properties is obtained for reasonable values of the parameters of the 4D gauge theory, if we apply our picture to a gauge theory of intermediate energy scale ($m_{\phi} \sim 10^{13}\,$GeV). Curiously, the inflaton mass is lower than $f$ ($\gtrsim M_P$) by several orders. This is due to the tiny SUSY breaking factor, $\beta \ll1$. $m_{\phi}$ and $f$ are related by 
\beq
m_{\phi}\sim g^2\beta f,
\eeq
(see $(\ref{infmass})$). Adding a small number of matter multiplets will not affect the conclusion significantly. \\
\indent We note that the region of observable inflation lies in the range $|\phi|>M_P$ in Fig. 2. Many of the inflation models proposed in the past \cite{Linde:1983gd, Freese:1990rb} posses this property. However their perturbative computation may not be trusted, because quantum gravity and other non-perturbative effects may upset the results. Arkani-Hamed et al.\cite{ArkaniHamed:2003wu} have pointed out that in higher-dimensional gauge theories no higher dimensional operators containing quantum gravity and other non-perturbative effects may be generated due to the gauge invariance in higher dimensions. Our results of inflaton potential using the region of large $|\phi|$ can thus be trusted.\\ \indent We may take a gauge theory in dimensions more than five. Then the Higgs and inflaton may be identified with different component of $A_M$; they are cousins rather than a single field. It is an interesting attempt to try to build a realistic model of Higgs and inflaton in this view.\\
\indent An alternative to the present approach is to identify the inflaton as the extra space components of the metric $g_{MN}$ instead of that of gauge field. In the context of string, this corresponds to considering closed string instead of open string. There have been some works in this direction \cite{radion-inflaton}.
 %%%%%%%%%%%%%%%%%%% Acknowledgements %%%%%%%%%%%%%%%%%%%%
\section{Acknowledgments}
  It is our pleasure to thank Nobuhito Maru for a valuable discussion at many stages of this work. T. I. wishes to thank Yuji Tachikawa for directing his interest to inflation models. We wish to thank Shinji Mukohyama and Chia - Min Lin for valuable discussions. We also wish to thank Pei-Ming Ho and Wen-Yu Wen for their kind hospitality and discussions during our visit to Taiwan University. This work is supported partially by the grants for scientific research of the Ministry of Education and Science, Kiban A, 18204024, Kiban B, 16340040 and Kiban C, 21540278 and by a Chuo University Riko-ken grant.

%%%%%%%%%%%%%%%%%%%%%%% reference %%%%%%%%%%%%%%%%%%%%%%%%


\begin{thebibliography}{100}

%\cite{Hatanaka:1998yp}
\bibitem{Hatanaka:1998yp}
  H.~Hatanaka, T.~Inami and C.~S.~Lim,
  %``The gauge hierarchy problem and higher dimensional gauge theories,''
  Mod.\ Phys.\ Lett.\  A {\bf 13}, 2601 (1998);\\
  %%CITATION = MPLAE,A13,2601;%%
%\cite{Antoniadis:1990ew}
  I.~Antoniadis,
  %``A Possible new dimension at a few TeV,''
  Phys.\ Lett.\  B {\bf 246}, 377 (1990).
  %%CITATION = PHLTA,B246,377;%%%\cite{ArkaniHamed:2003wu}
%\cite{ArkaniHamed:2003wu}
\bibitem{ArkaniHamed:2003wu}
  N.~Arkani-Hamed, H.~C.~Cheng, P.~Creminelli and L.~Randall,
  %``Extranatural inflation,''
  Phys.\ Rev.\ Lett.\  {\bf 90}, 221302 (2003).
  %%CITATION = PRLTA,90,221302;%%
%\cite{Kaplan:2003aj}
\bibitem{Kaplan:2003aj}
  D.~E.~Kaplan and N.~J.~Weiner,
  %``Little inflatons and gauge inflation,''
  JCAP {\bf 0402}, 005 (2004).
  %%CITATION = JCAPA,0402,005;%%
%\cite{radion-inflaton}

%\cite{Scherk:1978ta}
\bibitem{Scherk:1978ta}
  J.~Scherk and J.~H.~Schwarz,
  %``Spontaneous Breaking Of Supersymmetry Through Dimensional Reduction,''
  Phys.\ Lett.\  B {\bf 82}, 60 (1979).
  %%CITATION = PHLTA,B82,60;%%  
%\cite{Pomarol:1998sd}
\bibitem{Pomarol:1998sd}
  A.~Pomarol and M.~Quiros,
  %``The standard model from extra dimensions,''
  Phys.\ Lett.\  B {\bf 438}, 255 (1998);\\
  %%CITATION = PHLTA,B438,255;%%
    %\cite{SchmidtHoberg:2005yy}
  K.~Schmidt-Hoberg, DESY-THESIS-2005-009
    %``Supersymmetry breaking in five and six space-time dimensions,''
  %%CITATION = DESY-THESIS-2005-009;%%

   %\cite{Delgado:1998qr}
  \bibitem{Delgado:1998qr}
  A.~Delgado, A.~Pomarol and M.~Quiros,
  %``Supersymmetry and electroweak breaking from extra dimensions at the
  %TeV-scale,''
  Phys.\ Rev.\  D {\bf 60}, 095008 (1999);\\
  %%CITATION = PHRVA,D60,095008;%%
  %\cite{Takenaga:2003dp}
  K.~Takenaga,
  %``Effect of bare mass on the Hosotani mechanism,''
  Phys.\ Lett.\  B {\bf 570}, 244 (2003);\\
  %%CITATION = PHLTA,B570,244;%%
%\cite{Haba:2004xa}
  N.~Haba, K.~Takenaga and T.~Yamashita,
  %``Correct effective potential of supersymmetric Yang-Mills theory on M**4  x
  %S**1,''
  Phys.\ Rev.\  D {\bf 71}, 025006 (2005).
  %%CITATION = PHRVA,D71,025006;%%

 %\cite{Lyth:1998xn}
\bibitem{Lyth:1998xn}
  D.~H.~Lyth and A.~Riotto,
  %``Particle physics models of inflation and the cosmological density
  %perturbation,''
  Phys.\ Rept.\  {\bf 314}, 1 (1999).
  %%CITATION = PRPLC,314,1;%%
 
%\cite{Komatsu:2008hk}
\bibitem{Komatsu:2008hk}
  E.~Komatsu {\it et al.}  [WMAP Collaboration],
  %``Five-Year Wilkinson Microwave Anisotropy Probe (WMAP\altaffilmark 1 )
  %Observations:Cosmological Interpretation,''
  Astrophys.\ J.\ Suppl.\  {\bf 180}, 330 (2009).
  %%CITATION = APJSA,180,330;%%%\cite{Appelquist:1983vs}
\bibitem{Appelquist:1983vs}
  T.~Appelquist and A.~Chodos,
  %``The Quantum Dynamics Of Kaluza-Klein Theories,''
  Phys.\ Rev.\  D {\bf 28}, 772 (1983).
  %%CITATION = PHRVA,D28,772;%%

%\cite{Linde:1983gd}
\bibitem{Linde:1983gd}
  A.~D.~Linde,
  %``Chaotic Inflation,''
  Phys.\ Lett.\  B {\bf 129}, 177 (1983).
  %%CITATION = PHLTA,B129,177;%%
   %\cite{Freese:1990rb}
 \bibitem{Freese:1990rb}
   K.~Freese, J.~A.~Frieman and A.~V.~Olinto,
  %``Natural inflation with pseudo - Nambu-Goldstone bosons,''
  Phys.\ Rev.\ Lett.\  {\bf 65}, 3233 (1990).
  %%CITATION = PRLTA,65,3233;%%

\bibitem{radion-inflaton}
  N.~Arkani-Hamed, S.~Dimopoulos, N.~Kaloper and J.~March-Russell,
  %``Rapid asymmetric inflation and early cosmology in theories with
  %sub-millimeter dimensions,''
  Nucl.\ Phys.\  B {\bf 567}, 189 (2000);\\
  %%CITATION = NUPHA,B567,189;%% 
  C.~Csaki, M.~Graesser and J.~Terning,
  %``Late inflation and the moduli problem of sub-millimeter dimensions,''
  Phys.\ Lett.\  B {\bf 456}, 16 (1999);\\
  %%CITATION = PHLTA,B456,16;%%
  A.~Mazumdar, R.~N.~Mohapatra and A.~Perez-Lorenzana,
  %``Radion cosmology in theories with universal extra dimensions,''
  JCAP {\bf 0406}, 004 (2004).
  %%CITATION = JCAPA,0406,004;%%  

\end{thebibliography}
\end{document}